\documentclass{iopart}

\usepackage{fleqn}

\usepackage{ifthen}
\usepackage{ifpdf}

\usepackage{latexsym}
\usepackage{amssymb}
\usepackage{bm}

\ifpdf
\usepackage{graphicx}
\usepackage{epstopdf}
\else
\usepackage{graphicx}
\usepackage{epsfig}
\fi

\newcommand{\const}{\mbox{const}}

\newcommand{\im}{\mbox{Im}}

\newcommand{\tbox}[1]{\mbox{\tiny #1}}

\newcommand{\be}[1]{\begin{eqnarray}\ifthenelse{#1=-1}{\nonumber}{\ifthenelse{#1=0}{}{\label{e#1}}}}
\newcommand{\ee}{\end{eqnarray}}

\newcommand{\hide}[1]{}

\newcommand{\mpg}[2][\hsize]{\begin{minipage}[b]{#1}{#2}\end{minipage}}
\newcommand{\putgraph}[2][width=\hsize]{\includegraphics[#1]{#2}}

\begin{document}

\title[Mesoscopic Conductance]
{Rate of energy absorption by a closed ballistic ring}

\author{Doron Cohen$^{1}$, Tsampikos Kottos$^{2,3}$ and Holger Schanz$^{3,4}$}

\address{
$^{1}${Department of Physics, Ben-Gurion University, Beer-Sheva 84105, Israel} \\
$^{2}${Department of Physics, Wesleyan University, Middletown, Connecticut 06459-0155, USA } \\
$^{3}${Max-Planck-Institut f\"or Dynamik und Selbstorganisation G\"ottingen, Germany} \\
$^{4}${Institut f\"ur Nichtlineare Dynamik, Universit\"at G\"ottingen, Germany}
}


\begin{abstract}
We make a distinction between the spectroscopic and the mesoscopic
conductance of a closed ring. We show that the latter is not simply related
to the Landauer conductance of the corresponding open system. 
A new ingredient in the theory is related to the non-universal structure 
of the perturbation matrix which is generic for quantum chaotic systems.
These structures may created bottlenecks that suppress the diffusion 
in energy space, and hence the rate of energy absorption.
The resulting effect is not merely quantitative: 
For a ring-dot system we find that a smaller Landauer conductance 
implies a smaller spectroscopic conductance, 
while the mesoscopic conductance increases.
Our considerations open the way towards a realistic theory 
of dissipation in closed mesoscopic ballistic devices.
\end{abstract}

\section{Introduction}

When a physical system is subjected to an external perturbation 
it can absorb energy from the driving source (Fig.~1). 
The rate of this absorption depends crucially on the internal dynamics. 
Here we are concerned with a mesoscopic electronic system.
Mesoscopic means that the electrons must be treated as quantum mechanical 
particles whose wavelength is small compared to the classical dimensions. 
In such circumstances, which assume long coherence time,   
it is important whether or not the resulting electron 
dynamics would be integrable or chaotic or diffusive 
if approximated classically, and the {\em shape} of the device 
becomes relevant. The fingerprints of such non-universal (semiclassical) 
effects have been found in numerous experiments 
with {\em open} mesoscopic systems. It is our objective to extend 
this idea into the realm of {\em closed} mesoscopic systems, 
in the context of (semi-linear) response theory\footnote
{The response of chaotic systems to weak driving
is ``linear" in the classical treatment.  
We look for a quantum mechanical related departure
from linear response theory. This should be contrasted 
with the traditional studies of  ``quantum chaos" models
(such as the `quantum kicked rotator')
where in the absence of driving the system is integrable(!)
and consequently the response is manifestly non-linear
both classically and quantum mechanically.}. \\

{\bf Main observation:}
In this paper we expose a new ingredient in the 
theory of energy absorption by closed mesoscopic 
driven systems. The main idea is that there are 
circumstance in which the rate of absorption depends 
on the possibility to make {\em long sequences of transitions}.
The possibility to make a connected sequence 
of transitions between energy levels is greatly affected 
by structures in the energy landscape of the device.
These structures are the fingerprint of the ``shape" of the device, and more
generally they are implied by semiclassical considerations.  In the quantum
chaos literature such structures are termed non-universal so as to distinguish
them from the universal fluctuations which are described by random matrix
theory. In the context of energy absorption the most important non-universal
effect is the presence of ``bottlenecks'' in energy space where the couplings
between levels are small. The rate of energy absorption can be greatly reduced
by the presence of such bottlenecks. In order to take the effect of these
structures into account we have to go beyond the conventional framework of
linear response theory (LRT). \\


{\bf Main application:}
Closed mesoscopic rings are of great interest
\cite{rings,debye,IS,IS1,loc,G1,G2,kamenev,orsay}. 
For such devices the relation between the conductance 
and the internal dynamics is understood 
much less~\cite{kbf} than for open systems. 
This is not surprising if one realizes 
that the theoretical analysis of $G$ relies 
on completely different concepts in the {\em open} 
and in the {\em closed} case. 
For an open systems the Landauer formula expresses 
the conductance $G_{L}$ in terms of the scattering matrix 
of the device. This is a very convenient starting point 
for a subsequent analysis since it is not necessary 
to account for the dissipation of energy explicitly.  
By contrast, for closed systems the mechanism for dissipation 
is an issue.  Conventionally one assumes a weak coupling 
to an external bath 
which leads to a steady state but preserves approximately 
the main features of the internal dynamics.

First measurements of the conductance of closed 
mesoscopic rings have been reported more than 
a decade ago~\cite{orsay}.  
In a typical experiment a collection of mesoscopic rings 
are driven by a time dependent magnetic flux $\Phi(t)$ 
which creates an electro-motive-force (EMF) ${-\dot{\Phi}}$ 
in each ring. Assuming that Ohm's law applies, the induced current 
is ${I=-G\dot{\Phi} }$ and consequently the rate 
of energy absorption is given by Joule's law as 
\be{1000}
\dot{\mathcal{W}} 
\ \ \equiv \ \  \mbox{Rate of energy absorption} 
\ \ = \ \ G\,\dot{\Phi}^2 
\ee
where $G$ is called the conductance\footnote
{The terminology of this paper, and in particular 
our notion of ``conductance" are the same as in the 
theoretical review \cite{kamenev} and in the experimental work \cite{orsay}.}.
One should be very careful with the terminology here.  
We neither consider ``two terminal measurement" 
of the conductance nor ``four terminal 
measurement". One may say that closed ring  
is a ``zero terminal" device (no leads). 
So when we say ``conductance" we relate 
to the coefficient $G$ in Eq.(\ref{e1000}).      
In practice $G$ is deduced form 
a measurement of the magnetic 
susceptibility $\chi(\omega)$. 
If we have a large collections of rings 
then $G$ should be identified as 
the $\omega\rightarrow0$ limit of 
$\im[\chi(\omega)]/\omega$ divided 
by the number of rings. 
We are aware that some of the people 
in the condensed matter community avoid  
the use of the term ``conductance"  
for a zero terminal device. 
There are no leads attached, so a transport 
measurement in the sense of the Landauer geometry 
is not applicable. However, we are not aware 
of a better name for $G$ of Eq.(\ref{e1000}). 
We think that the terminology issue  
is mainly a matter of taste.   
Namely, if the ring were 1~meter in diameter,  
rotating between the magnets of a commercial 
generator, no one would be bothered by 
calling $G$ ``conductance", so why not to adopt 
the same terminology in the nano scale?

In {\em typical circumstances}, 
which we define more precisely later on, 
the energy absorption process is dominated by
Fermi-Golden-Rule (FGR) transitions, 
and the strong dynamical localization 
effect~\cite{loc} is irrelevant. 
{\em We shall always assume this FGR regime
and neglect other dissipation mechanisms} 
such as Landau-Zener transitions \cite{wilk}
or Debye relaxation \cite{debye}.  
The FGR picture with some extra assumptions 
(that we would like to challenge) is the basis for LRT. 
It leads to the Kubo formula, 
which is a major tool in many fields 
of theoretical physics and Chemistry.  
Diagrammatic techniques of calculating conductance 
are based on this formula. \\

{\bf Past works:} 
In the case of {\em diffusive rings} the Kubo formula leads 
to the Drude formula for $G$.  A major challenge for past 
studies was to calculate and to measure 
weak localization corrections to the Drude result, 
taking into account the level statistics  
and the type of occupation. For a review see~\cite{kamenev}. 
It should be clear that these corrections 
do not change the leading order Kubo-Drude result. \\

{\bf Present work:} 
In the case of {\em ballistic rings} we would like 
to argue that even if the FGR picture of energy absorption 
applies,  still there are circumstance where the Kubo formula, 
and hence also the Drude result, fail even as a rough approximation.
It should be clear that our theory is not in contradiction with LRT.
Rather it goes beyond LRT, and reduces to LRT in the appropriate limit.  
The failure of the Kubo formula is related to non-universal 
features of the energy landscape which are implied by having 
a {\em mean free path larger than the size of the system}.
To some extent our theory is inspired by ideas from percolation theory: 
but {\em the percolation is in energy space rather than in real space}. \\

{\bf Scope:} 
In this paper we analyze an example of a single mode device, 
while in a follow-up work~\cite{bls} we analyze an example of 
a multi-mode ring. The single mode example of this paper 
(unlike the multi mode example of~\cite{bls}) possibly seems 
somewhat artificial. Its advantage is its simplicity. 
Our interest in this paper is to clarify the main idea of the 
conductance calculation with one simple prototype example, 
rather than exploring the full range of possibilities.       
By now another non-trivial extension of the theory 
has been worked out in Ref.~\cite{slr}, where the absorption of small 
metallic grains is calculated. The latter reference has suggested  
to describe the outcome of the theory as ``semilinear response".  \\

{\bf Outline of this paper:} 
This paper consists of two parts. 
In the first part (sections~2-9) 
we present the general theoretical 
consideration. In particular:
\begin{itemize}
\item We define   
the notion of ``conductance" 
in the context of closed systems.
This leads to our distinction 
between mesoscopic and spectroscopic conductance. 
\item We state our main results regarding 
the conductance of a single mode ballistic device,   
and discuss their experimental significance. 
\item We review the FGR picture, 
and explain the emergence of diffusion 
in energy space, and the associated 
dissipation effect.    
\item We clarify the main ingredient 
in our theory, which is the calculation 
of the coarse grained diffusion 
\item We derive a formula for the 
mesoscopic conductance assuming 
a modulated energy landscape. 
\end{itemize}
In the second part we focus 
on a specific example: 
a single mode ring-dot model.  
We express the spectroscopic 
and the mesoscopic  conductance as a function 
of the averaged Landauer conductance 
of the corresponding open system.   
We explore the dependence of the result 
on the level broadening $\Gamma$. 
Finally we summarize some key observations
and point out again the limitations 
on the validity of our results.

\section{Mesoscopic versus Spectroscopic conductance}

For the understanding of this paper it is crucial 
to make a distinction between the spectroscopic conductance 
$G_{\tbox{spec}}$ and the mesoscopic conductance $G_{\tbox{meso}}$. 
Both are defined as the dissipation coefficient  
which appears in Eq.(\ref{e1000}), but they relate 
to different circumstances. The bottom line is 
that $G_{\tbox{spec}}$ is the measured value 
of $G$ for small EMF, while $G_{\tbox{meso}}$ is 
the measured value of $G$ for large EMF. In later 
sections we shall explain that $G_{\tbox{spec}}$
can be calculated using the traditional version 
of the Kubo formula, while $G_{\tbox{meso}}$ 
requires a different recipe. Thus it should 
be clear that our theory does not contradict LRT 
but rather covers a regime where LRT is no longer valid.

In order to explain the physical picture of energy 
absorption we refer to the block diagram 
of Fig.~1. On the one hand the driving source 
induce transitions between energy levels of the system, 
leading to an absorption of energy with some 
rate  $\dot{\mathcal{W}}$.
On the other hand the system can release
energy to some bath (phonons or the surrounding),  
which leads to a ``heat flow" with some 
rate $\dot{\mathcal{Q}}$. 
As the system heats up a steady state is reached 
once $\dot{\mathcal{Q}}=\dot{\mathcal{W}}$.

It is essential to realize that the levels of the system 
are effectively ``broadened" due to the non-adiabaticity 
of the driving~\cite{pmc} or due to the interaction 
with the noisy environment~\cite{IS}. Later we quantify 
this effect by introducing the level broadening 
parameter $\Gamma$. One should not make a confusion between 
\be{-1} 
\Gamma &=& \mbox{level broadening parameter in the theory} 
\\ \nonumber
\gamma_{\tbox{rlx}} &=& \mbox{relaxation rate towards equilibrium} 
\ee
The above distinction is somewhat analogous 
to the notions of $1/T_2$ and $1/T_1$ in NMR studies. 
The parameter $\Gamma$ is essential in order  
to analyze the induced FGR transitions between 
levels. Therefore it will appear explicitly     
in the theoretical derivations. We are going to argue 
that the rate of energy absorption $\dot{\mathcal{W}}$ 
is sensitive to $\Gamma$.  
The relaxation parameter  $\gamma_{\tbox{rlx}}$ plays 
a different role: it is responsible for achieving 
a steady state. Furthermore, as explained below 
it is an essential input in order to predict  
whether the value of the conductance 
is $G=G_{\tbox{spec}}$ or $G=G_{\tbox{meso}}$.

{\bf Algebraic average:} 
Let us assume that we have a large collection 
of similar rings with possibly a broad thermal 
population of the energy levels.  
If the energy landscape is not 
uniform, it is evident that some rings 
are likely to absorb energy, while others 
are not, depending on whether the initial 
level is strongly coupled to its neighboring 
levels or not. Accordingly the {\em initial} rate 
of energy absorption (per ring) 
is obtained by a simple {\em algebraic average}.
This algebraic average reflects the 
statistical nature of the preparation and 
has nothing to do with the nature of the dynamics.

{\bf Slow down:} 
When we measure $G$ we are not interested 
in the transient behavior but rather 
in the long time behavior. The possible scenarios 
are illustrated in Fig.~1. One possibility 
is to have small $\gamma_{\tbox{rlx}}$. 
In such case the rate of absorption slows down: 
in the ``long run" the rate of energy absorption
is limited by the bottlenecks. 
[The reader can easily make here an analogy   
with the dynamics of traffic flow]. 
We would like to mention that the above scenario 
is further discussed in the context 
of a later work~\cite{slr}. 
There the reader can find an actual numerical 
simulation that demonstrates the transient 
from an initial large rate of absorption 
to a much slower long time rate of absorption.

{\bf Mesoscopic circumstances:}
The above scenario features a crossover 
from a large absorption rate to a slower 
absorption rate. 
The transient is characterized 
by a time scale that we call $t_{\tbox{stbl}}$.
In a later section we shall determine 
this time scale and its dependence 
on the EMF: The larger $\dot{\Phi}$
the smaller $t_{\tbox{stbl}}$. 
The condition for observing the 
slow-down is obviously 
\be{1001}
t_{\tbox{stbl}}(\dot{\Phi}) \ \ \ll \ \ \gamma_{\tbox{rlx}}^{-1}  
\ee   
The rate of absorption for $t \gg t_{\tbox{stbl}}$   
depends on the possibility to make {\em long sequences of transitions}. 
Hence the value of $G$ is smaller compared 
with the initial anticipation.  
This value is what we call~$G_{\tbox{meso}}$.

{\bf Spectroscopic circumstances:}
Thus the theory for $G_{\tbox{meso}}$ becomes 
relevant whenever the relaxation process 
is not efficient enough to mask the intrinsic dynamics 
of the device. Optionally one may say that 
the EMF should not be too small. What happens 
if $\dot{\Phi}$ becomes small, 
such that Eq.(\ref{e1001}) breaks down? 
In such case the relaxation process  
re-initiated the initial distribution
before the slow-down shows up. Consequently 
the drop in the rate of absorption is avoided.  
The value of $G$ in the latter circumstances 
is what we call~$G_{\tbox{spec}}$.

\section{Description of the model system}

For sake of analysis we would like to define 
the {\em simplest model} where our general idea can 
be demonstrated. As explained in the introduction 
there is a large class of systems, where the 
perturbation matrix ($\mathcal{I}_{nm}$) 
is {\em structured}. In particular this is the case 
with ballistic devices. The simplest would 
be to consider a one channel device driven by EMF, 
hence $\mathcal{I}_{nm}$ are the matrix elements 
of the current operator. In order 
to have some weak scattering mechanism 
we consider ring which is weakly connected 
to a big dot region. A particle that moves 
inside the ring has some small probability to 
enter into the dot region, where its velocity 
is randomized.  We are going to argue that  
the long-time energy absorption process 
is not determined by a simple algebraic average 
over~$|\mathcal{I}_{nm}|^2$, but rather 
involves a non-trivial coarse graining procedure. 
We would like to figure out what 
is the conductance of such ring-dot device as 
a function of the ring-dot coupling.

To be specific we analyze the model  
which is illustrated in Fig.~2a, 
where the dot is modeled as a big chaotic network. 
The ring states mix with the dense set of the dot
states, and resonances are formed. Large $\mathcal{I}_{nm}$ 
matrix elements are found only between states within resonances, 
thus leading to a structured band profile (see Fig.~3b). 
We assume DC driving: this means that the driving frequency 
is much smaller compared with the energy scales  
that characterize the structures of $\mathcal{I}_{nm}$. 
We will show that the off-resonance regions form bottlenecks 
for the long-time energy absorption. 
As a result we get $G_{\tbox{meso}} \ll G_{\tbox{spec}}$. 
It should be clear that if $\mathcal{I}_{nm}$ had no structure 
we would get $G_{\tbox{meso}} = G_{\tbox{spec}}$.

In the presented analysis {\em dynamical localization}  
effect is ignored. This mechanism also leads to 
suppression of energy absorption, but it involves 
much longer time scales. Moreover, it is extremely 
sensitive to decoherence, and to any temporal 
irregularity of the driving. Thus in 
typical realistic experimental circumstances 
the suppression of diffusion due to `bottlenecks' 
is much more likely than the suppression of diffusion  
due to a dynamical localization effect.

\section{Main results and their experimental significance}

Before we dive into the derivations, we would 
like to give an overview of the main results, 
and to discuss their experimental significance.  
We also exploit the opportunity to comment on 
the results of a follow-up work~\cite{bls} 
where multi-mode chaotic rings have been analyzed.

If the device were opened as in Fig.~2c  
we could have asked what is its the Landauer 
conductance. It is convenient to {\em characterize}  
the device by the energy averaged Landauer 
conductance. The averaging is over the relevant 
energy range around ${E \sim E_F}$. The energy   
window of interest is assumed to be classically small 
but quantum mechanically large (many levels):   
\be{0}
g_{cl} = \overline{ g_L(E) }
\ee 
We use $0<g<1$ rather than $G$ in order 
to indicate that the conductance of the 
is measured in units of $e^2/(2\pi\hbar)$. 
The subscript of $g_{cl}$ further 
imply that the energy averaged Landauer conductance 
yields the {\em classical} transmission of the device.

From a theoretical standpoint we can 
ask what is the ``conductance" 
of the same device if it is integrated   
in a closed circuit as in Fig.~2b.
If we could (hypothetically) ignore the quantum 
interference within the ring, then we would 
get (section~9) the ``Drude" result 
\be{0}
G_{\tbox{Drude}} = 
\frac{e^2}{2\pi\hbar}  
\left[\frac{g_{cl}}{1-g_{cl}}\right]
\ee
which diverges in the ${g_{cl} \rightarrow 1}$ limit.
Indeed we are going to argue that a similar 
result is obtained for the spectroscopic conductance. 
We say ``similar" rather than "identical" because 
quantum mechanics sets an upper bound to $G$.

If the environmentally induced relaxation 
is weak, we argue that the spectroscopic 
result is wrong. Then we have to calculate 
the mesoscopic conductance using 
a recipe that we are going to develop 
in sections~5-8. This leads to 
\be{100}
G_{\tbox{meso}} =  \frac{e^2}{2\pi\hbar}  \ (1-g_{cl})^2 g_{cl}
\ee
One observes that in the limit ${g_{cl} \rightarrow 1}$  
(weak coupling of the ring with the dot) 
the mesoscopic conductance goes to zero.
This should be contrasted with the behavior
of the Drude result.

Measurements of the conductance of closed  
mesoscopic rings have been performed already~10 years 
ago~\cite{orsay}. In a practical experiment 
a large array of rings is fabricated.
The conductance measurement can be achieved  
via coupling to a highly sensitive electromagnetic 
superconducting micro-resonator. In such setup the EMF 
is realized by creating a current through 
a ``wire" that spirals on top of the array,  
and the conductance of the rings is determined via 
their influence on the electrical circuit. 
Another possibility is to extract the conductance  
from the rate of Joule heating. The later can 
be deduced from a temperature difference measurement 
assuming that the thermal conductance is known.

Ballistic devices are state-of-the-art in mesoscopic experiments.   
Moreover we believe that molecular size devices with 
closed ``ring geometry" are going to be of 
great interest in the near future. It is likely that 
the dynamics is such devices would be of ballistic nature. 
Namely, it is likely that the mean free path in such 
rings would be larger compared with their perimeter. \\

{\bf Single-mode rings:} 
The results in the present paper apply 
to single mode devices with ring-dot geometry. 
Such a device can be realized in practice. 
Furthermore, by incorporating a gate, one can 
control the ring-dot coupling. Hence such geometry 
looks optimal for an experimental test of the theory:  
The conductance can be measured as a function 
of the coupling, and at least the qualitative 
agreement with Eq.(\ref{e100}) can be tested. \\

{\bf Multi-mode rings:} 
In a follow-up work~\cite{bls} we analyze
the mesoscopic conductance of multi-mode chaotic 
rings. For example one can picture such a device   
as a wide ring with a small gate-controlled deformation. 
Thus it is possible to test the theory by  
measuring~$G$ as a function of the gate voltage 
determined $g_T$.    
It is explained in~\cite{bls} that  
also in the case of a multi mode ring 
the eigenfunctions are {\em non-ergodic} 
for $(1-g_T) \ll 1$. Consequently absorption is 
suppressed due to not having {\em ''connected sequences 
of transitions"}. It should be noted however that 
in general there is no simple structure of resonances,  
in contrast to the 1~mode case that we are going to analyze.  
Therefore an accurate estimate of the mesoscopic 
conductance requires a more elaborated resistor 
network analogy.

\section{The Kubo-Einstein formula for the diffusion}

The purpose of this section is to review 
a well known expression for the diffusion 
coefficient $D$. If we were talking about 
diffusion in real space, then it is well known 
from any statistical mechanics textbook  
that $D$ is equal to the integral over 
the velocity-velocity correlation function. 
This is known as the Einstein formula. 
Similarly, the diffusion coefficient in energy space  
is related to the integral over the 
current-current correlation function. 
This is regarded by some authors~\cite{wilk} 
as a particular version of the Kubo formula,     
because with some extra assumptions it leads 
to the ``popular" version of the 
Kubo formula for the conductance~$G$.

We define the one-particle current operator $\mathcal{I}$  
in the conventional way as the symmetrized version 
of ${e\hat{v}\delta(\hat{x}-x_0)}$, with $\hat{v}=\hat{p}/\mathsf{m}$, 
where $\mathsf{m}$ and $e$ are the mass and the charge of a spinless electron, 
and $x=x_0$ is a section through which the current is measured.  
In the later numerical analysis it was convenient 
to re-define $\mathcal{I} := \int dx_0\, a(x_0) \mathcal{I}$ 
where $a(x_0)$ is a wide weight function whose integral  
over the ring obeys  $\oint a(x)dx = 1$.  
The current-current correlation function 
for an electron with energy $E$ is 
\be{0}
C_E(\tau) = \langle \mathcal{I}(\tau) \mathcal{I}(0) \rangle 
\ee
It is customary to symmetrize this function, but 
this is not essential since we later use it 
within a $d\tau$ integral that goes 
from~$-\infty$ to~$\infty$.
The power spectrum of
the fluctuations is defined as the Fourier transform of the correlation
function $C_E(\tau)$.  In the quantum case it is related to the matrix
elements of the current operator as follows~\cite{mario}:
\be{5}
\tilde{C}_E(\omega)  
\ = \
\left[
\sum_{n (\ne m) }
|{\cal I}_{nm}|^2
\,\, 2\pi\delta_{\Gamma}\left(\omega-\frac{E_m{-}E_n}{\hbar}\right)
\right]_{E_m \sim E}
\ee
where the smoothing parameter $\Gamma$ is introduced 
because it is required in a later stage.

Having defined the current-current 
correlation function, we can write 
the Kubo-Einstein formula for 
the EMF-induced diffusion in energy space:  
\be{312}
D_E = \frac{1}{2} \int_{-\infty}^{\infty} C_E(\tau)d \tau  
\times {\dot {\Phi}}^2
\ee 
The units of $D_E$ are such 
that $\delta E^2= 2D_E t$ 
for a local spreading of a wavepacket.
In appendix~A we summarize the 
derivation of this formula and 
also write the  associated 
diffusion equation that describes the 
evolution of an arbitrary distribution $\rho(E)$.  
An optional way to write the Kubo-Einstein formula is 
\be{313}
D_E = \frac{1}{2} \tilde{C}_E(\omega{=}0) \times {\dot {\Phi}}^2
\ee 
We immediately see that in the quantum 
mechanical case the Kubo-Einstein formula 
is ill-defined unless we specify the parameter $\Gamma$.
Note that for $\Gamma=0$ we get formally $D_E=0$. 
The physics behind this formula becomes more transparent 
if we adopt the Fermi-Golden rule 
picture as discussed in the next section.

\section{The Fermi Golden Rule Picture}

The Hamiltonian of the ring system 
in the adiabatic basis is  
\be{0}
\mathcal{H} \mapsto E_n\delta_{nm} + W_{nm} 
\ee
where 
\be{0}
W_{nm} = i\dot{\Phi} \frac{\hbar\mathcal{I}_{nm}}{E_n{-}E_m}
\ee
is the perturbation matrix. If the EMF is non-zero 
then there are transitions between levels. 
Both Linear Response Theory, and also our 
extended theory of mesoscopic conductance  
assume that the rate of transition    
from an initial level ($m$) 
to some other level ($n$) is determined 
by Fermi Golden Rule (FGR): 
\be{0}
w_{nm} \ \  = \ \  \frac{2\pi}{\hbar} 
\delta_{\Gamma}(E_n-E_m) \ |W_{nm}|^2
\ee
Note that upon summation over all transitions 
to all the levels~$n$, the delta function 
is replaced by the smoothed density of states~$\varrho(E)$.

In the {\em adiabatic regime}, the level 
broadening $\Gamma$ is smaller compared 
with the mean level spacing $\Delta$. 
In such case the FGR mechanism 
can be neglected ($w_{nm}\sim0$), 
and the leading dissipation 
mechanism, depending on the effectiveness 
of the environmental relaxation process, 
is either the Landau-Zener mechanism~\cite{wilk}, 
or the Debye relaxation mechanism~\cite{debye}. 
In the present work we assume 
that $\Gamma$ is much larger than $\Delta$,  
but much smaller compared with any  
other semiclassical energy scale.
This implies that FGR transitions 
are the dominant mechanism for 
diffusion in energy space.

Due to the FGR transitions there 
is a diffusion in energy space. 
The local diffusion rate is: 
\be{314}
D_E 
\ \ = \ \ 
\frac{1}{2} \sum_m (E_n-E_m)^2 \ w_{nm} 
\ \ = \ \ 
\pi\hbar \varrho(E) \overline{|\mathcal{I}_{nm}|^2} \times {\dot {\Phi}}^2
\ee
The first expression in Eq.(\ref{e314}) 
is just as in the standard analysis 
of a random walk problems. 
Upon substitution of the FGR expression  
it leads to Eq.(\ref{e313}). 
The second expression in Eq.(\ref{e314})
is just a loose way to re-write Eq.(\ref{e313}). 
The dependence on~$\Gamma$ is implicit.

\section{Coarse grained diffusion}

The probability distribution in energy 
space obeys a diffusion equation (see appendix~A).
If we change form the variable $E$ to 
the variable~$n$ the diffusion equation 
takes a simpler form: 
\be{201}
\frac{\partial \rho(n)}{\partial t} \ = \
\frac{\partial}{\partial n}
\left(D_n \frac{\partial}{\partial n}
\rho(n)\right)
\ee
where $D_n=\varrho^2D_E$.
Our main motivation to re-write the 
diffusion equation in the~$n$ variable 
is to suggest the analogy with the 
familiar problem of random walk 
on a lattice where~$n$ is re-interpreted 
as a site index. In the standard 
textbook discussion $D_n$ is uniform all 
over space. {\em But what happens if $D_n$ 
has some microscopic modulation?} 
We would like to argue that in such case 
the coarse grained diffusion coefficient  
is given by an harmonic 
average over the local $D_n$. Namely, 
\be{401}
D = \langle \langle D_n  \rangle\rangle \ \ \equiv \ \  
\left[ \lim_{N\rightarrow\infty} \frac{1}{N} \sum_{n}^{N} 
D_n^{-1} \right]^{-1} 
\ee
This is like calculating the 
resistivity of a {\em random network} 
of resistors in one dimension. 
The summation can be re-interpreted  
as the addition of resistors in series.  
This analogy is further developed 
in a subsequent work~\cite{slr}. 
Below we derive this result using 
the diffusion picture language. 
It is important to realize that the validity  
of the harmonic average recipe Eq.(\ref{e401}) 
is limited. It is assumed that $D_n$ has  
a smooth modulation as a function of with~$n$.  
Otherwise~$n$ cannot be treated as a continuous 
variable, and one should use the more elaborated 
``random network" scheme of calculation.

In order to keep consistency of notations 
we turn back to use~$E$ as the 
diffusion space variable, as in appendix~A. 
For simplicity we assume that locally 
the smoothed density of states is 
constant. Hence the diffusion equation 
is Eq.(\ref{e201}) with $n$ replaced by $E$.   
It can be regarded as a continuity equation 
\be{0}
\frac{\partial \rho(E)}{\partial t} \ = \
-\frac{\partial}{\partial E}J_E
\ee
where the probability current (probability 
transported per unit time) is given by Fick's law
\be{0}
J_E = - D_E \frac{d\rho(E)}{dE}  
\ee
In order to determine the coarse 
grained diffusion we assume a
steady state distribution $\rho(E)$ 
that supports a current $J_E=\const$.  
It follows that 
\be{0}
\rho(E_2)-\rho(E_1) = - \int_{E_1}^{E_2} (J_E/D_{E}) dE
\ee
This implies that the coarse-grained
diffusion coefficient is given by 
\be{-1} 
D^{-1} = -\frac{1}{J_E} \frac{\rho(E_2){-}\rho(E_1)}{E_2-E_1} 
= \frac{1}{E_2{-}E_1} \int_{E_1}^{E_2} D_{E}^{-1} dE
\ee
where $E_2-E_1$ is the coarse graining scale.
Hence we get the desired result: the coarse 
grained diffusion coefficient~$D$ is obtained 
by harmonic average over the modulated~$D_E$. 
A concise way to write this result for 
the coarse grained diffusion is  
\be{0}\label{cgdc} 
D = \langle\langle D_E \rangle\rangle
= \left[ \overline{ 1/D_E } \right]^{-1} 
\ee 
where the indicated average 
in the r.h.s. is over the energy $E$.  
This observation is going to be the corner 
stone in the analysis of the mesoscopic conductance.
Needless to say that the harmonic average recipe
implies vanishing $D$ if $D_E$ has bottlenecks.

\section{The diffusion-dissipation relation}

There is a simple relation between the diffusion 
and the rate of energy absorption. For a derivation 
of this relation see appendix~A. In the case 
of a low temperature Fermi occupation it takes 
the form 
\be{1010}
\dot{\mathcal{W}}
\ \ = \ \ 
\varrho(E_F) \langle\langle D_E \rangle\rangle
\ee
where $E_F$ is the Fermi energy. 
By substitution of Eq.(\ref{e313}) 
and comparing with Eq.(\ref{e1000}) 
we deduce the following expression 
for the conductance:
\be{6}
G \ = \ \varrho(E_F) \times \frac{1}{2} 
\langle \langle \tilde{C}_{E}(\omega{=}0) \rangle \rangle
\ee
From the discussion in section~2 it is implied  
that in the case of the spectroscopic conductance $G_{\tbox{spec}}$, 
the energy averaging ${\langle \langle .. \rangle \rangle}$
should be {\em algebraic}.  In the case of diffusive rings 
it is customary to replace the (algebraic) energy averaging 
by disorder averaging.  
In contrast to that from the discussion in section~7 it is implied  
that in the case of the mesoscopic conductance $G_{\tbox{meso}}$, 
the energy averaging ${\langle \langle .. \rangle \rangle}$
should be {\em harmonic}. To be more precise, the harmonic 
average applies if $D_E$ has smooth modulation 
as a function of $E$. This is going to be the case 
with our simple example. In more complicated circumstances 
(e.g. multimode chaotic rings) the coarse graining 
should be done using a somewhat more elaborated resistor 
network analogy~\cite{slr,bls}.

In the case of $G_{\tbox{spec}}$ calculation, 
Eq.(\ref{e6}) is just the conventional Kubo formula. 
If the energy landscape is uniform then the distinction 
between $G_{\tbox{spec}}$ and $G_{\tbox{meso}}$ is 
not important. But if $\mathcal{I}_{nm}$
is sparse or structured then $G_{\tbox{meso}}$ 
might be much smaller compared with $G_{\tbox{spec}}$.

{\bf Sensitivity to level broadening:}
The parameter $\Gamma$ is an implicit input 
for the calculation. As discussed in section~6 
we assume that there is effectively 
a quasi-continuum ($\Gamma\gg\Delta$).  
The spectroscopic conductance $G_{\tbox{spec}}$ 
is not very sensitive to $\Gamma$. 
For example, in the case of diffusive rings 
the weak localization corrections are 
of order $(\Gamma/\Delta)^{-1}$.   
In contrast to that $G_{\tbox{meso}}$ 
is extremely sensitive to $\Gamma$. 
By increasing $\Gamma$ we can enhance 
the diffusion by several orders 
of magnitudes. As common in the 
traditional LRT treatment \cite{kamenev} 
also here $\Gamma$ is going to be a free parameter in the theory.

\section{Semiclassical considerations}

If the current-current correlation function 
is calculated classically, one observes 
a and very slow monotonic dependence on $E$. 
Since the practical interest is in a classically 
small energy window, this dependence can be ignored. 
For a one dimensional ring whose 
classical transmission is $g_{cl}$ 
the correlation function $C_E(\tau)$ 
decays exponentially. The intensity 
of the fluctuations $\tilde{C}_{E}(\omega{=}0)$ 
that appears in Eq.(\ref{e6}) equals 
the area under this classical 
correlation function leading to~\cite{kbf}: 
\begin{equation}
\label{Gclas}
G_{\tbox{Drude}} 
= \frac{e^2}{2\pi\hbar} \left( \frac{g_{cl}}{1-g_{cl}} \right)
\end{equation}
This is simply the Drude result being written 
in terms of $g_{cl}$ instead 
in terms of the mean free path $\ell \approx L_0/(1-g_{cl})$, 
where $L_0$ is the length of the ring.  
We note that if we had neglected the multiple reflections 
within the ring (which is formally justified in 
case of an open system) we would get $g_{cl}$ instead 
of $g_{cl}/(1-g_{cl})$ in agreement 
with the Landauer formula.

The ``classical" result is obtained also within  
the framework of a semiclassical Green function 
calculation~\cite{pmt} that employs a diagonal approximation. 
Such calculation assumes an {\em algebraic average} 
over the energy or if applicable, over realizations of disorder. 
We recall that algebraic average is justified 
for the purpose of calculating the spectroscopic conductance.  
Still the spectroscopic conductance cannot diverge 
in the limit ${g_{cl}\rightarrow 1}$. We shall discuss 
the upper bound on $G_{\tbox{spec}}$ in section~12.

As stated above in the classical analysis $C_E(\tau)$  
and hence $D_E$ are essentially independent of~$E$. 
But upon quantization $D_E$ might have a modulation 
on a classically small energy scale $\Delta_0\propto\hbar$ 
which is still much larger compared with the mean 
level spacing $\Delta$. The energy scale $\Delta_0$ 
reflects the appearance of a structured band profile 
and it is the new ingredient in our analysis.
The time to explore this energy scale is 
\be{0}
t_{\tbox{stbl}} 
= \frac{\Delta_0^2}{\langle\langle D_E \rangle\rangle} 
\ \ \propto \ \ \dot{\Phi}^{-2}
\ee
Assuming that $D_E$ has indeed a ``microscopic" dependence 
on $E$, and assuming that Eq.(\ref{e1001}) is satisfied, 
we have argued that the long time energy absorption is determined 
by the mesoscopic conductance, which involves an harmonic 
average.  Consequently the mesoscopic conductance 
is typically smaller than the spectroscopic conductance, 
and does not correspond to the classical (Drude) result!


\section{Preliminary analysis of the model system}

We turn now to a detailed description of our model (Fig.~2a). 
The dot region is modeled as a chaotic network which 
is weakly coupled to a ring. The network is attached 
with two ``legs" in order to destroy odd-even symmetries 
of the eigenstates. Each coupling vertex is described 
by a $3\times 3$ symmetric orthogonal matrix 
(the ``splitter" of Ref.\cite{splitter}): 
\be{-1}
\bm{S}=
\left(\begin{array}{ccc}
\frac{1}{2}(1-\sqrt{1-2c^{2}}) & \frac{1}{2}(1+\sqrt{1-2c^{2}}) & c\\
\frac{1}{2}(1+\sqrt{1-2c^{2}}) & \frac{1}{2}(1-\sqrt{1-2c^{2}}) &-c\\
c & -c & \sqrt{1-2c^{2}}
\end{array}\right)
\ee
where the coupling parameter is ${0<c<1}$.  
Once the chaotic network is integrated into the ring 
one can regard it as a ``black box" which is   
characterized by a $2\times 2$ scattering matrix. 
This $2\times 2$ scattering matrix is characterized 
by the average Landauer conductance $g_{cl}$.  
Hence the coupling between the network and the
ring is quantified by the dimensionless 
parameter $1-g_{cl} \approx c^2$ (see Fig.~4b), 
which we assume to be much smaller than~1.

The length of the ring is $L_0$ while 
the total length of all the bond is $L$.
Resonances are formed because of the coupling 
of the ring states to the network states. Hence we have
\be{7}
\Delta &=& \hbar v_E \frac{\pi}{L} = \mbox{mean level spacing} 
\\
\Delta_0 &=& \hbar v_E \frac{\pi}{L_0} = \mbox{distance between resonances} 
\\ 
\label{gamma0}
\Gamma_0 &=& \frac{1}{2\pi} (1-g_{cl}) \Delta_0 = \mbox{width of resonances} 
\ee 
where $v_E=(2E/\mathsf{m})^{1/2}$. Having defined the dimensionless parameters 
of the model system ($L/L_0$ and $g_{cl}$) as well as the relevant 
energy scales $(\Delta,\Delta_0,\Gamma_0)$, we are fully equipped to turn to 
the calculation of the conductance.


A key object in the analysis is the weight $q(E_n)$ of each state in the ring
region.  We write the wavefunction on the ring 
as $\psi^n(x\in\mbox{ring}) = \mathsf{A}_n \sin(\varphi_n+k_nx)$ and define
\be{11}
q(E_n) = \frac{L}{2}|\mathsf{A}_n|^2
\approx \frac{L}{L_0} \sum_r 
|\langle r | n \rangle |^2
\ee
where $r$ are the ring states in the absence 
of coupling ($c=0$). If we could assume 
ergodicity of the wavefunctions then $A_n=(2/L)^{1/2}$ 
and we would get $q(E)=1$.  
But if the coupling becomes weak ($1-g_{cl} \ll 1$) 
then ergodicity does not hold (Fig.~4a) and we have $\overline{q}=1$ 
only upon averaging over energy.  
Using perturbation theory we get 
as the simplest approximation that $q(E)$ 
is a sum of Lorentzian: 
\be{12}
q(E) \ \approx \ 
\frac{\Delta_0}{\pi} \sum_{r} 
\frac{(\Gamma_0/2)}{(\Gamma_0/2)^2+(E-\varepsilon_r)^2}
\ee
where the resonance energies $\varepsilon_r$ have spacing $\Delta_0$.
It is important to notice that the above expression 
does not reflect that     
\be{13}
\mbox{maximum}[q(E_n)] = \frac{L}{L_0} = \frac{\Delta_0}{\Delta} 
\ee
The maximum corresponds to the extreme case of having 
all the probability of the wavefunction inside the ring 
such that $A_n=(2/L_0)^{1/2}$. This situation is attained 
if we decrease the coupling so as to have no mixing 
of ring states with dot states ($\Gamma_0 \sim \Delta$).  
In the following discussion we assume that the latter 
(trivial possibility) is not the case. 
Hence we observe from Eq.(\ref{e12}) 
that ${q(E) \approx  (1-g_{cl})^{-1}}$ 
on resonances (where $E\sim\varepsilon_r$),  
while ${q(E) \approx (1-g_{cl})}$ 
off resonances (where $|E-\varepsilon_r| \sim \Delta_0  \gg \Gamma_0 $).  
The algebraic energy average over $q^2(E)$ is dominated  
by the peaks within resonance regions, 
whereas the harmonic average is dominated by the bottleneck  
off-resonance (valley) regions. The relative size 
of the resonance regions is $\Gamma_0/\Delta_0 \approx (1-g_{cl})$. 
Therefore we get 
\be{14}
\ \overline{q^2}  &\approx& (1-g_{cl})^{-1}
\\ \label{e15}
\ [\overline{1/q^2}]^{-1}  &\approx& (1-g_{cl})^2
\ee

\section{Calculation of the conductance for the model system}


Given the eigenfunctions of the network 
the matrix elements of ${\cal I}$ are
\be{-1}
{\cal I}_{nm} \approx  
-i\frac{ev_{\tbox{F}}}{2} \mathsf{A}_n\mathsf{A}_m 
\sin([\varphi_n-\varphi_m]+[k_n-k_m]x_0)\,.
\ee
Without loss of generality we set $x_0=0$ and find 
\be{16}
\frac{2\pi\hbar}{\Delta}
|I_{nm}|^2 
= 
e^2\frac{v_{\tbox{F}}}{L} 
\  q(E_n) \ q(E_m) \times g_{\varphi}
\ \ \ \ \ \ \ \ 
\mbox{for ${|E_n-E_m| \ll \Delta_0}$}
\ee
where $0<g_{\varphi}<1$ is defined as the average value 
of $2|\sin(\varphi_n-\varphi_m)|^2$ for nearby levels  
If we change $x_0$ the correction to $g_{\varphi}$ 
is at most of order $((E_n-E_m)/\Delta_0)^2$ 
and hence can be neglected.
Now we can get an explicit result 
for $\tilde{C}_E(\omega)$ via Eq.(\ref{e5}) 
and hence for the conductance:
\be{17}
G \ = \  \frac{e^2}{2\pi\hbar} 
\langle\langle q(E)q_{\Gamma}(E+\hbar\omega)\rangle\rangle \times g_{\varphi}
\ee
where $q_{\Gamma}(E)$ is a smoothed version 
of $q(E)$ as implied by Eq.(\ref{e5}). 
Note that in the numerical analysis one  
should be careful to use a smoothing kernel 
that excludes the center element, as implied by  
the restriction~${m \ne n}$.

Using  Eq.(\ref{e17}) with Eq.(\ref{e14}) 
we get for the spectroscopic conductance 
\be{0}
G_{\tbox{spec}} = \left[\frac{e^2}{2\pi\hbar}\right]  
\  (1-g_{cl})^{-1} \ g_{\varphi}
\ee 
Comparing this to Eq.~(\ref{Gclas}) 
we deduce $g_{\varphi}=g_{cl}$. 
On the other hand using  Eq.(\ref{e17}) 
with Eq.(\ref{e15}) we get 
\be{0}
G_{\tbox{meso}} = \left[\frac{e^2}{2\pi\hbar}\right]  
\  (1-g_{cl})^2 \ g_{\varphi}
\ee 
which leads to Eq.(\ref{e100}) 
for the mesoscopic conductance. 
We can re-phrase the above procedure as follows:  
Given the well behaved ($\Gamma$~insensitive) 
spectroscopic conductance, 
the mesoscopic conductance is 
\be{2}
G_{\tbox{meso}} =   [(\overline{1/q^2})^{-1}/\overline{q^2}] \times G_{\tbox{spec}}
\ee
with $\Gamma$ implicit in the definition of $q$. 
Thus in order to calculate $G_{\tbox{meso}}$ we need 
to know both $G_{\tbox{spec}}$ and $q_{\Gamma}(E)$.  
For flat band ($q(E)=1$) and we get $G_{\tbox{meso}}=G_{\tbox{spec}}$, 
but in general $G_{\tbox{meso}} < G_{\tbox{spec}}$.


\section{Numerics}

Numerical tests have been done for the model system of Fig.~2a.  
We have chosen a network containing $25$~bonds of length~$\sim1$ 
such that $L \approx 25$. Our energy window was set around $k \sim 25000$.  
We computed $(k_n, \varphi_n, \mathsf{A}_n)$ 
and with these data we have done the analysis
described above.  In particular we have obtained $q(E)$ for various values of
the couplings (see Fig.~4a).  Then we have calculated both the algebraic and
the harmonic averages of $q^2(E)$.  From the results which are presented in
Fig.~4b we were able to deduce that the relation between these averages 
and $g_{cl}$ is as expected from the theoretical considerations.

In Fig.~5 we show that the numerical result   
for $G_{\tbox{spec}}$ and $G_{\tbox{meso}}$ 
is sensitive to the smoothing parameter $\Gamma$. 
Note that the displayed range $\Gamma$ exceeds 
the physically relevant region.  
In the adiabatic regime ($\Gamma \le \Delta$) 
the results are unstable numerically 
and of no significance.  Also very large values 
of $\Gamma$ are not physically significant.  

We observe in Fig.~5 that $G_{\tbox{spec}}$ cannot 
exceed the quantum bound, and becomes saturated  
in the limit $g_{cl}\to 1$. This saturation is 
attained for $(1-g_{cl})^{-1}>L/L_0$ 
as implied by the quantum border of Eq.(\ref{e13}):  
\be{0}
G_{\tbox{spec}} \Big|_{\tbox{maximum}} 
= \left[\frac{e^2}{2\pi\hbar}\right]  
\  \frac{L}{L_0}
\ee 
We have a second set of data (not displayed) 
for a similar network with $L/L_0$ 
larger by a factor~$3.5$. We have verified that 
the saturation value of $G_{spec}$ 
increases by the same factor.

\newpage

\section{Summary}

The original motivation for this work was to study 
the simplest model of a closed device where the FGR picture 
of transitions and the notion of conductance are meaningful.
In order to construct such a model quantum chaos considerations 
are essential. A ring with a simple scatterer is not good enough: 
Energy diffusion cannot be justified 
due to adiabaticity or strong localization effect~\cite{loc}.
Therefore we have replaced the scatterer by a complicated
network, which has allowed us to define circumstance such that 
the FGR picture would apply. Our initial inclination was to 
assume that in such circumstance we could use the Kubo formula
in order to calculate the rate of energy absorption.   
But then we have realized that the applicability 
of the FGR picture is not enough in order to get Kubo.  
This turned out to be an example for a system where 
the conductance depends crucially 
on the non-universal {\em structure} of the perturbation matrix.

In other systems with structured band profile 
a similar effect is expected,
although not necessarily as dramatic 
as in our simple model.  In particular we note 
a subsequent work~\cite{bls} where we calculate  
the conductance of multi mode ballistic rings.
Thus {\em a feature that looked like an anomaly 
of a single mode device, has turned out to be 
a general theme in the theory of energy absorption}. 
A further non-trivial extension of the theory 
has allowed also to calculate the absorption 
of small metallic grains that are irradiated  
by a low frequency noise source~\cite{slr}.

The model system that we have picked is of particular interest because it
adds a twist to the old discussion of the Landauer formula \cite{imry}. 
For an open system the two terminal conductance of a one-channel system
is less than unity.  For a closed (zero terminal) device we have 
to distinguish between spectroscopic and mesoscopic conductance.  
The former equals $g_{cl}/(1-g_{cl})$ and can be very large 
in the limit of weak scattering, while the latter equals $(1-g_{cl})^2 g_{cl}$ 
and goes to zero in this limit (disregarding higher-order corrections).  
More generally we can say
that within the validity limits of our assumptions,  
$g_{cl}$ is the upper limit for
the mesoscopic conductance of a single-mode system.  
There is a numerical indication~\cite{bls}, not yet conclusive,  
that also in the case of a multimode ring the mesoscopic conductance
is limited by the number of open modes.

The calculation of the conductance of closed devices 
is sensitive to the level broadening parameter $\Gamma$. 
Hence theoretical considerations that go well beyond 
the common Kubo formalism are essential. 
$\Gamma$ is determined by the non-adiabaticity 
of the driving~\cite{pmc} and/or by the 
surrounding environment~\cite{IS}. 
We have assumed in the present paper  
that $\Delta \ll \Gamma \ll \Delta_0$. 
Our results do not apply to the adiabatic regime
where other (rival) dissipation mechanisms apply  
(Landau-Zener~\cite{loc}, Debye~\cite{debye}).
On the other extreme, for very large driving rate (EMF) 
we may have a non-perturbative response~\cite{crs} 
that can invalidate the fluctuation-diffusion relation 
on which the calculation of $D_E$ has been based.

\newpage
\appendix
\section{The Kubo-Einstein and the diffusion-dissipation relations}

In this appendix we review the derivation of  
the Kubo-Einstein formula and the diffusion-dissipation  
relation following \cite{wilk,frc}. 
The diffusion in energy space is deduced from 
the relation 
\be{0}
\frac{d{\cal H}}{dt} 
\ \ = \ \ \frac{\partial{\cal H}}{\partial t} 
\ \ = \ \ \dot{\Phi} \times \frac{\partial{\cal H}}{\partial \Phi} 
\ \ = \ \ - \dot{\Phi} \times \mathcal{I}
\ee
This exact relation holds both classically 
and quantum mechanically. In the latter case 
we have to use Heisenberg picture. For simplicity of 
presentation we use a classical language and write 
\be{0}
E(t)-E(0) \ = \ - \dot{\Phi} \int_0^t {\cal I}(t') dt'
\ee
averaging over an initial microcanonical preparation we get 
\be{0}
\langle (E(t)-E(0))^2 \rangle \ = \ 
\dot{\Phi}^2 \int_0^t\int_0^t 
\langle {\cal I}(t') {\cal I}(t'') \rangle_E \ dt'dt''
\ee
Thus 
\be{0}
\delta E^2(t) \ = \ 2D_E t
\ee
where $D_E$ is given by Eq.(\ref{e312}). 
On long times one can argue \cite{jar} that the probability 
distribution $\rho(E)$ of the energy
should satisfies the following diffusion equation:
\be{0}
\frac{\partial \rho}{\partial t} \ = \
\frac{\partial}{\partial E}
\left(\varrho(E)D_E \frac{\partial}{\partial E}
\left(\frac{1}{\varrho(E)}\rho\right)\right)
\ee
where $\varrho(E)$ is the density of states.
The energy of the system is
$\langle {\cal H} \rangle=\int E \rho(E)dE$.
It follows that the rate of energy absorption is
\be{0}
\dot{{\cal W}} \ = \ 
\frac{d}{dt}\langle {\cal H} \rangle
= - \int_0^{\infty} dE \ \varrho(E) \ D_E
\ \frac{\partial}{\partial E}
\left(\frac{\rho(E)}{\varrho(E)}\right)
\ee
For a Fermi occupation $\rho(E)=\varrho(E)f(E)$ 
where $f(E)$ is the occupation function. 
At zero temperature we get Eq.(\ref{e1010}).

\newpage

\ \\ 

\ack

DC thanks Michael Wilkinson, Bernhard Mehlig, Yuval Gefen 
and Shmuel Fishman for intriguing discussions. 
The preparation of the expanded version of this paper 
has been inspired by the helpful suggestions of Miriam Blaauboer, 
and we thank Boris Shapiro and Markus B\"{u}ttiker for their 
additional comments. The research was supported by 
the Israel Science Foundation (grant No.11/02),
and by a grant from the GIF, the German-Israeli Foundation 
for Scientific Research and Development.



\newpage

\mpg{
\putgraph[width=0.7\hsize]{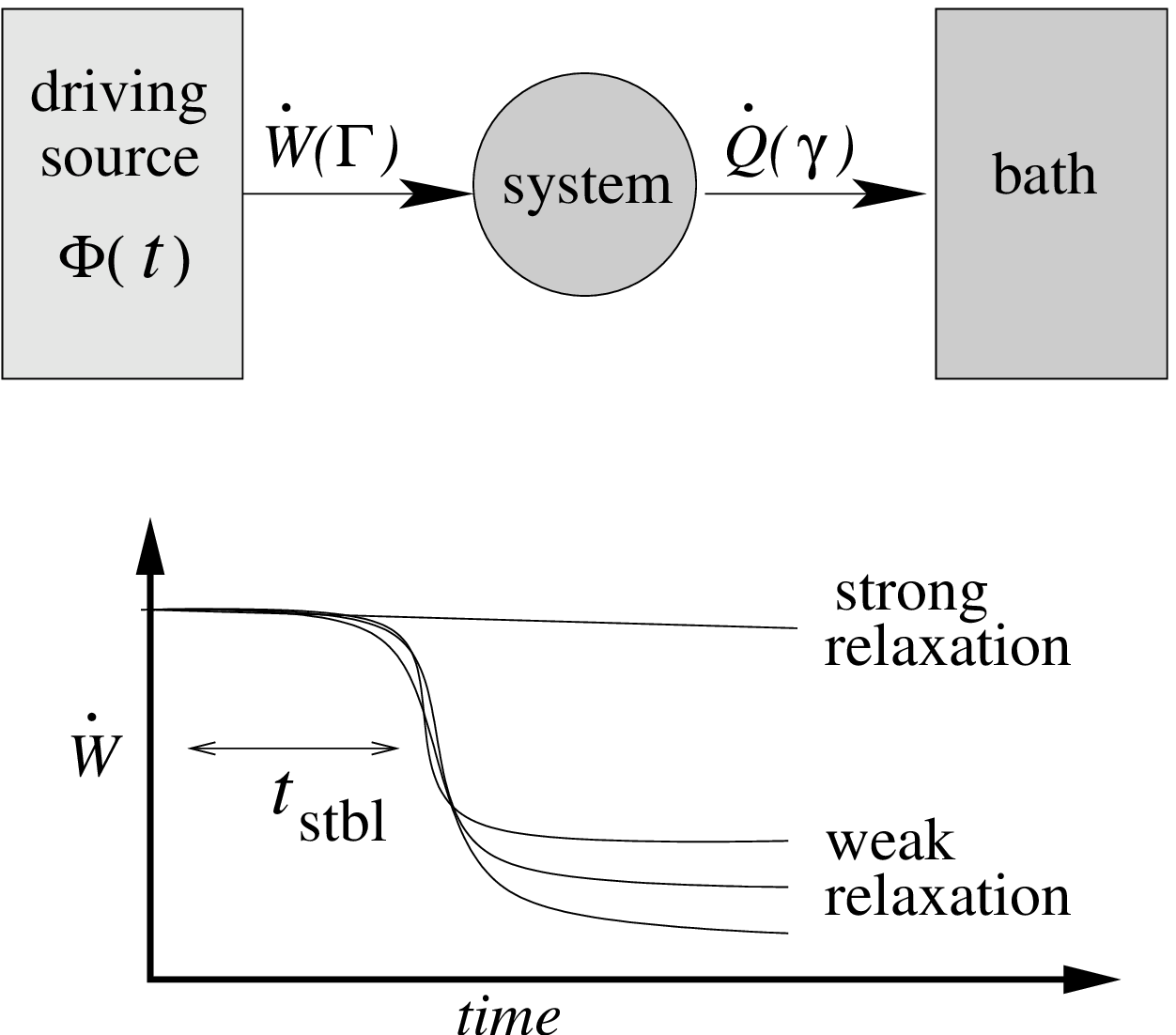}

{\footnotesize {\bf Fig.1}:
The phenomenology of energy absorption. 
In the upper panel we present a block 
diagram.  We highlight the rate of energy 
absorption $\dot{\mathcal{W}}$ which depends on the 
energy broadening parameter $\Gamma$. 
We also highlight the rate $\dot{\mathcal{Q}}$ 
of ``heat flow" to the bath,  
which depends on the energy 
relaxation rate $\gamma_{\tbox{rlx}}$. 
In the lower panel we plot how $\dot{W}$ 
depends on time. If $\gamma_{\tbox{rlx}}$ 
is small than there is a transient to a slower 
absorption rate that depends on $\Gamma$.}
}

\ \\ \ \\ \ \\

\mpg{
\putgraph[width=0.9\hsize]{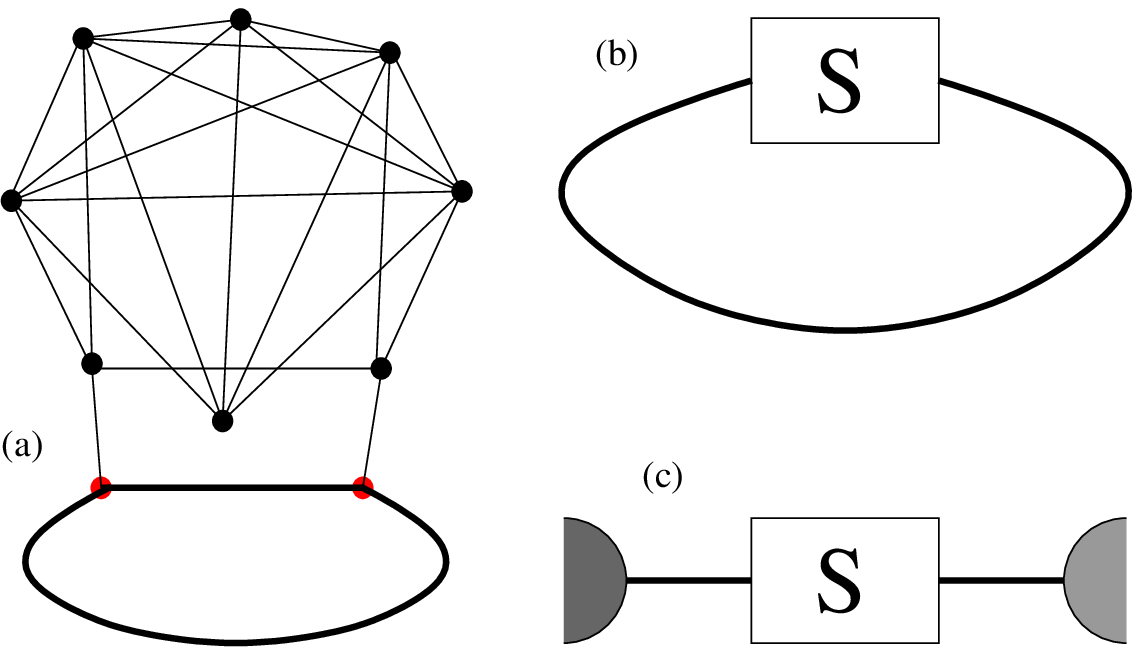}

{\footnotesize {\bf Fig.2}:
(a) The ring-dot network model. 
(b) Schematic representation. 
(c) The corresponding open system.}
}

\newpage

\mpg{
\putgraph[width=0.45\hsize]{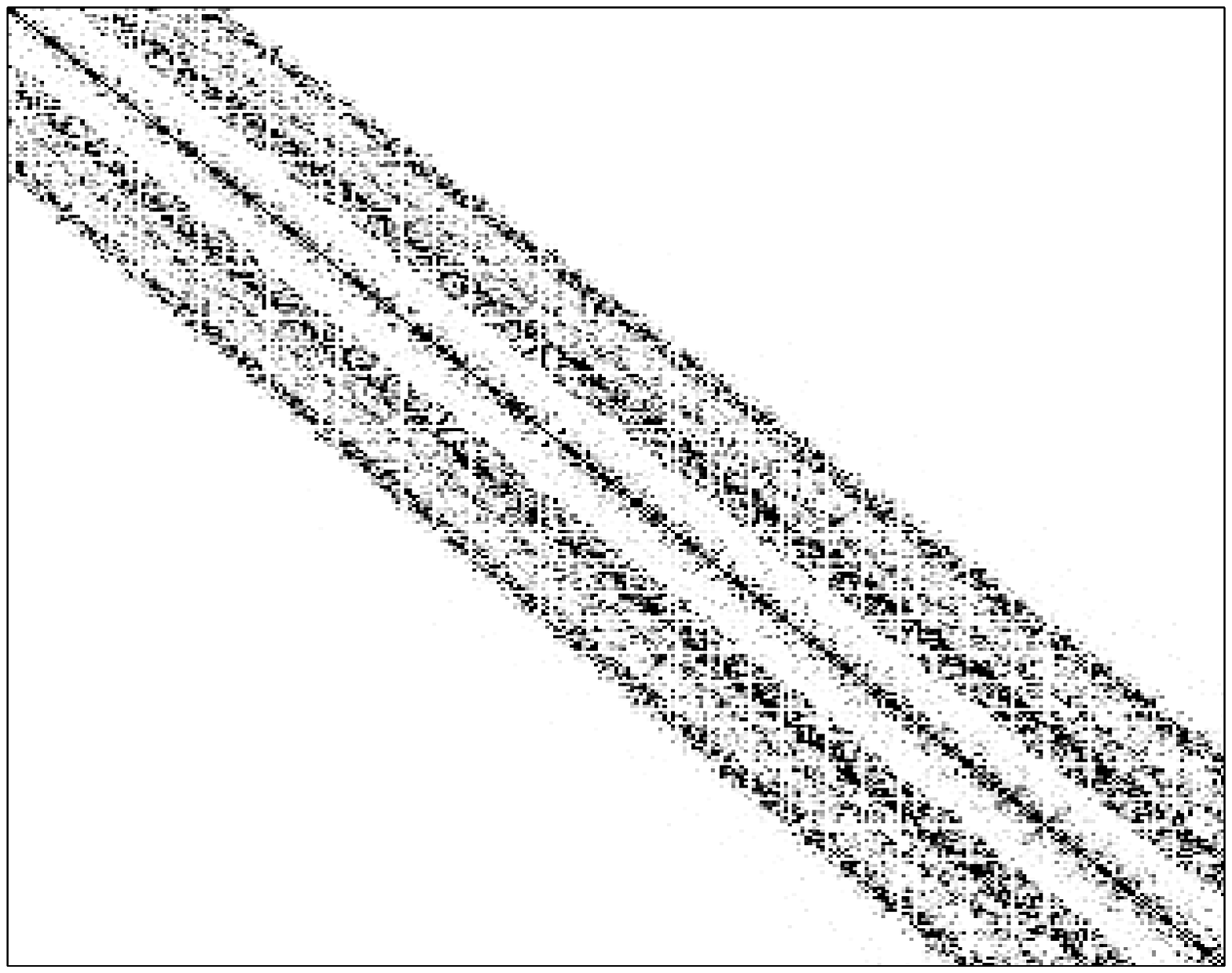}
\putgraph[width=0.45\hsize]{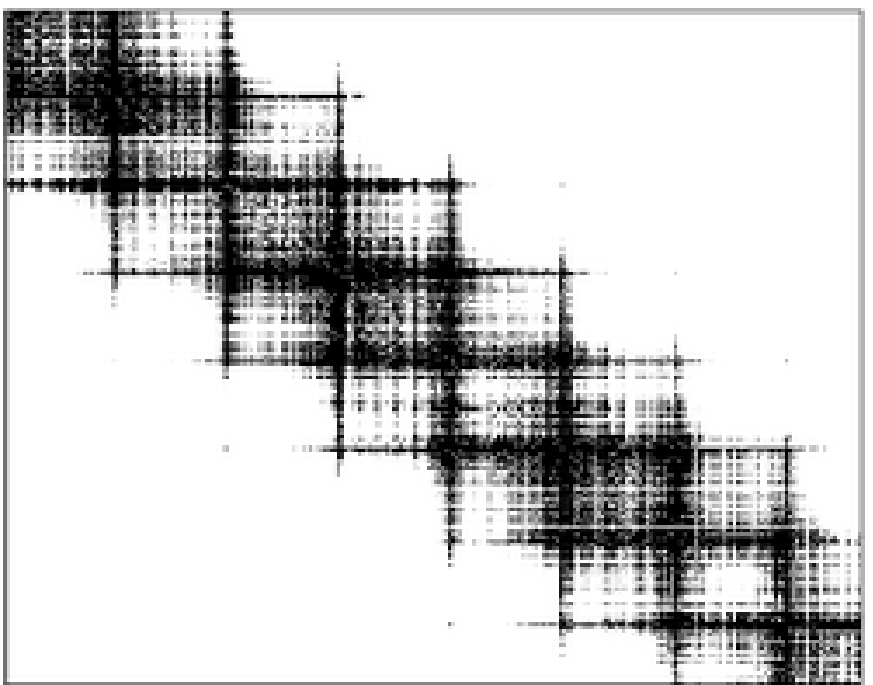}

{\footnotesize {\bf Fig.3}: 
Writing the perturbation in the basis 
of the unperturbed Hamiltonian we get 
a structured matrix: On the left we 
display a representative result for   
a quantized hard chaos system (taken form Ref.~\cite{lds}).  
On the right we display the result 
for the ring-dot network model of this paper.}  
}

\ \\  

\mpg{
\putgraph[width=0.46\hsize]{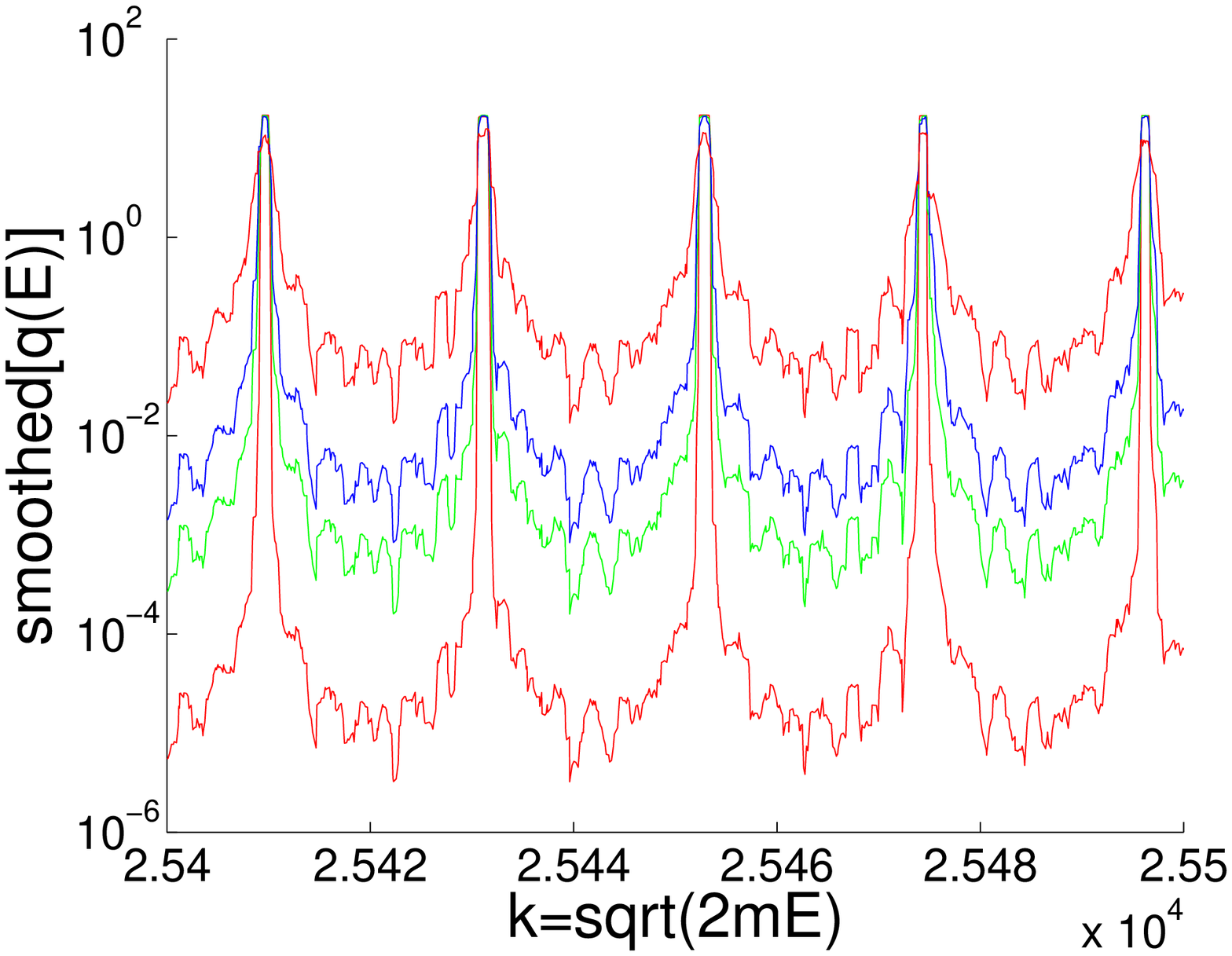}
\putgraph[width=0.46\hsize]{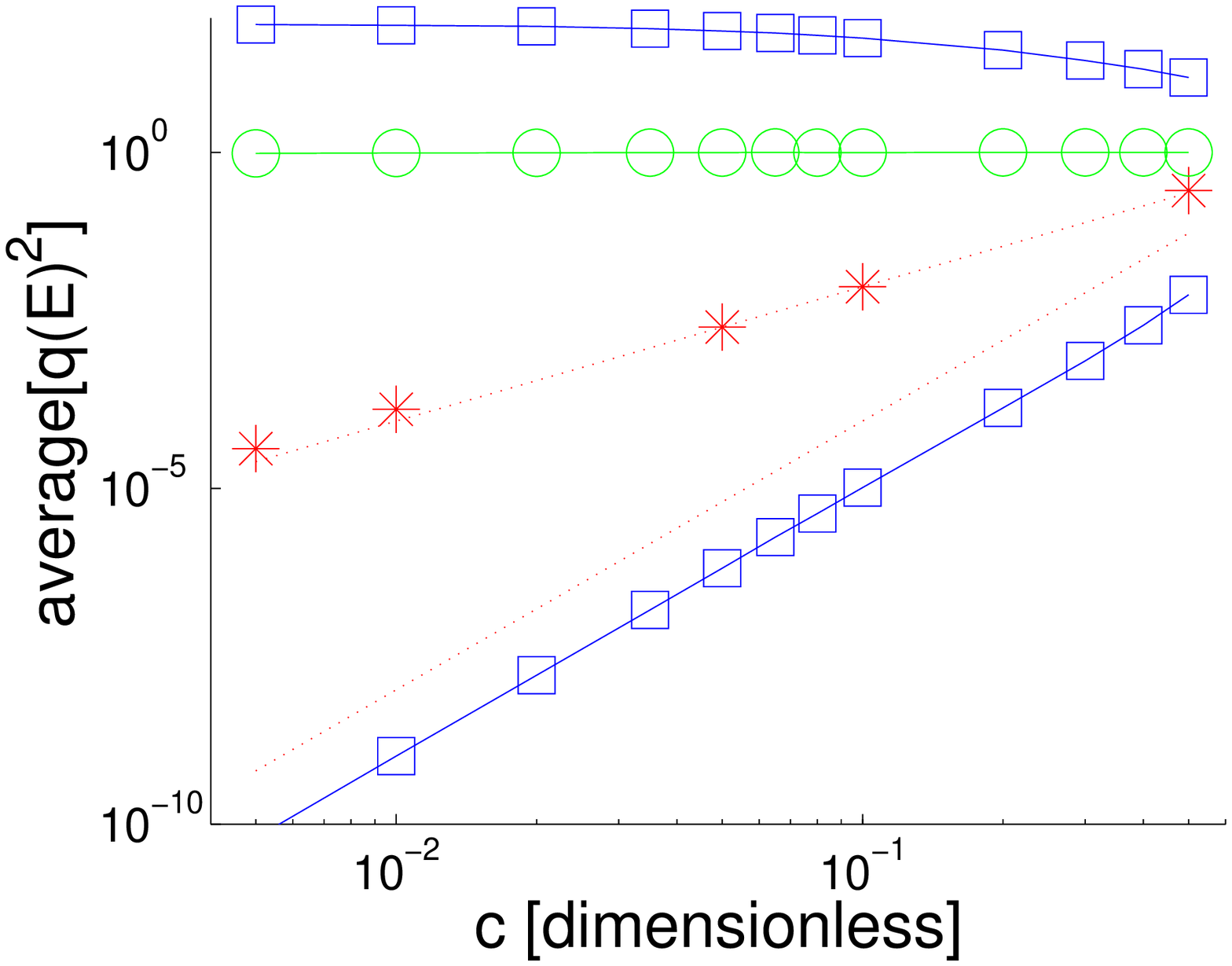}

{\footnotesize {\bf Fig.4:}  
Left panel: Smoothed $q(E_n)$ for $0.005<c<0.5$.
Right panel: The horizontal axis is $c$. 
The dotted red lines are $c^2$ and $c^4$. 
The averages $1/\overline{[1/q(E)^2]}$ (lower curve), 
and $\overline{[q(E)^2]}$ (upper curve), 
and $\overline{[q(E)]}$ (circles) are calculated. 
The coupling measure $1-g_{cl}$ (stars) is 
numerically obtained from the Landauer conductance.} 
}

\ \\

\mpg{
\putgraph[width=0.46\hsize]{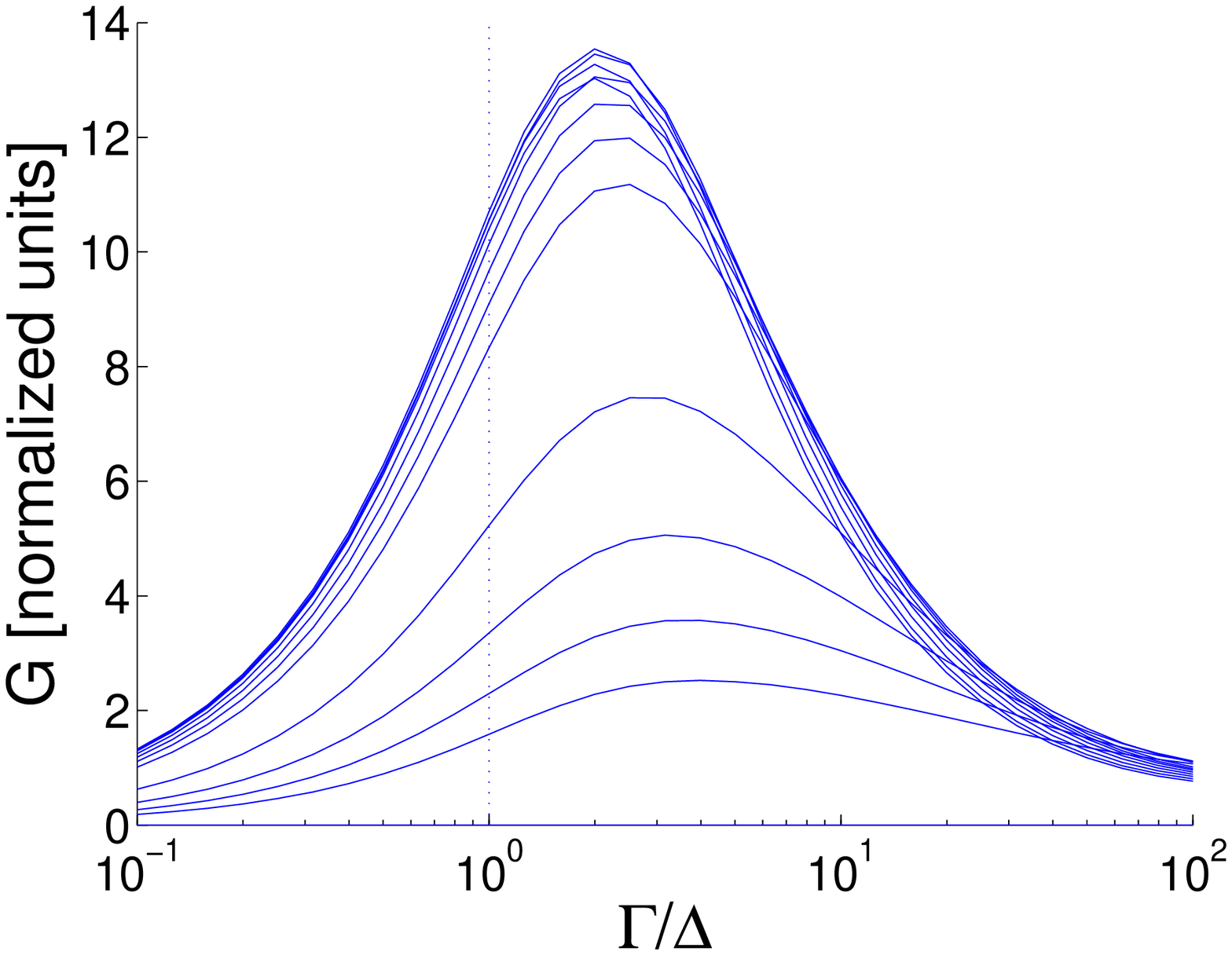}
\putgraph[width=0.46\hsize]{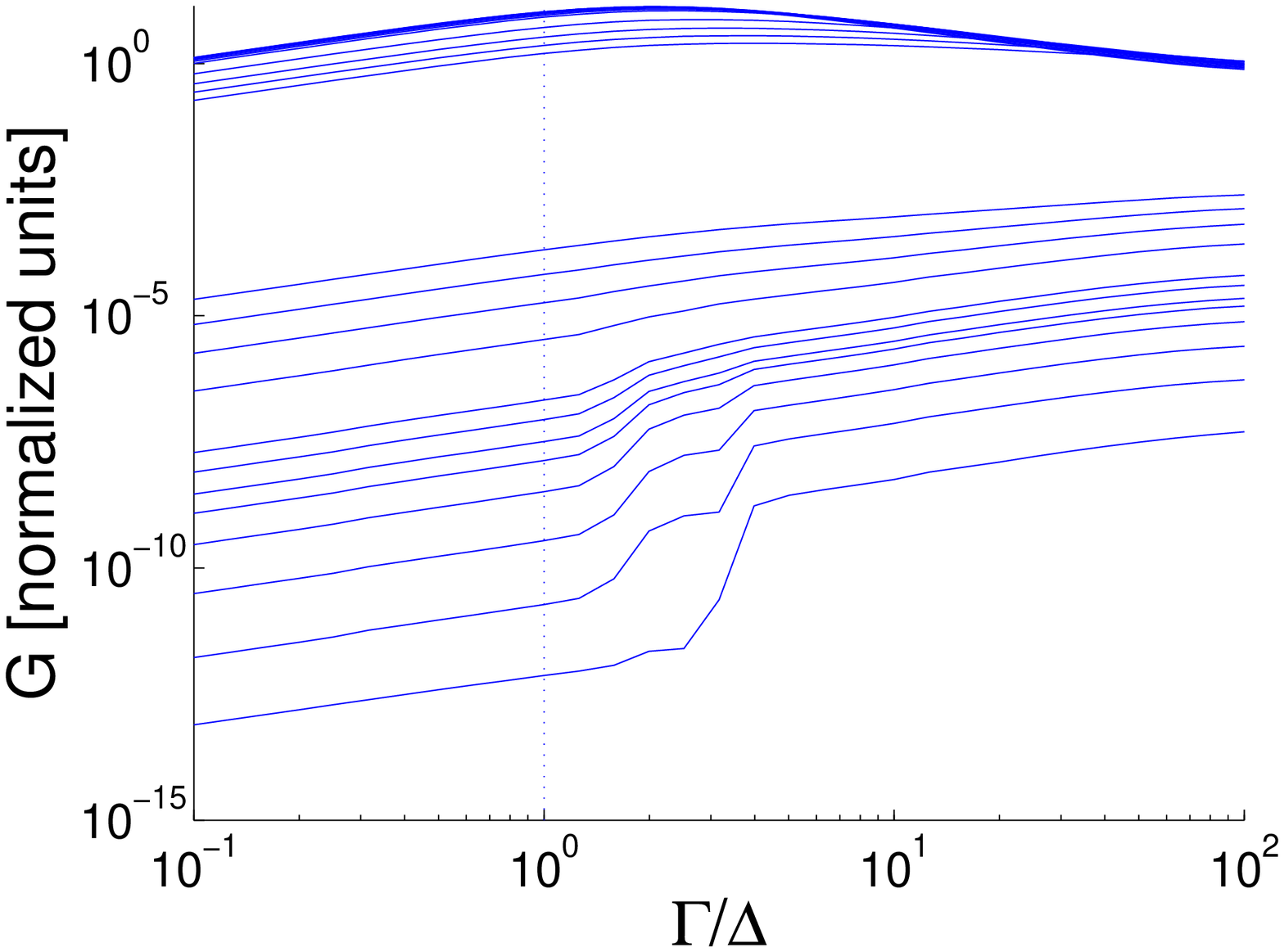}

{\footnotesize {\bf Fig.5:}  
The spectroscopic and mesoscopic conductances 
as a function of $\Gamma/\Delta$ 
for the various values of $c$. 
The former is resolved in the normal scale (left) 
while the latter is resolved in the log scale (right). 
The vertical dotted line indicates 
the $\Gamma$ value that has been assumed in Fig.~4b. } 
}

\end{document}